\documentclass[a4paper,11pt]{article}
\usepackage{pos} % for details on the use of the package, please see the JINST-author-manual
\usepackage{lineno}
\usepackage{array,multirow,graphicx}
\usepackage{bbm}
\usepackage{hyperref}
\usepackage[dvipsnames]{xcolor}
\usepackage{float}
\usepackage{hhline}
\usepackage{multicol}
\usepackage{graphicx}
\usepackage{soul}
\usepackage{csquotes}
\usepackage{tikz-feynman}
\tikzfeynmanset{/tikzfeynman/every gluon@@/.style={
    /tikz/draw=none,
    /tikz/decoration={name=none},
    /tikz/postaction={
      /tikz/draw,
      /tikz/decoration={coil,aspect=0,
        amplitude=1.8mm,segment length=4.5mm, post length=0.5pt % Shorten the final straight segment
      },
      /tikz/decorate=true,
    }
}}
\usepackage{subcaption}
\usepackage{booktabs}
\usepackage{physics}
\usepackage{siunitx}
\usepackage{soul}
\usepackage[normalem]{ulem}
\RequirePackage[noabbrev]{cleveref}
\crefname{equation}{eq.}{eqs.}
\RequirePackage[status=draft,inline,nomargin,marginclue]{fixme}
\fxusetheme{color}
\FXRegisterAuthor{fh}{afh}{FH}
\FXRegisterAuthor{mc}{amc}{\color{purple}MC}
\FXRegisterAuthor{vg}{avg}{\color{orange}VG}

\newcommand{\alphaem}{\alpha_{\mathrm{em}}}
\newcommand*\diff{\mathop{}\!\mathrm{d}}

\title{New parton distribution functions of the 
real photon}

\author*[a,b]{M. Chithirasreemadam}
\author[a,b]{V. Guzey}
\author[a,b]{F. Hekhorn}
\author[a,b]{I. Helenius}
\author[a,b]{H. Paukkunen}

\affiliation[a]{Department of Physics, University of Jyvaskyla, P.O. Box 35, FI-40014 University
of Jyvaskyla, Finland}
\affiliation[b]
{Helsinki Institute of Physics, University of Helsinki, P.O. Box 64, FI-00014 University of Helsinki, Finland}

\emailAdd{madhav.m.chithirasreemadam@jyu.fi}

\abstract{

Precise determination of the partonic structure of real photons has attracted renewed interest in view of ongoing studies of high-energy photon-induced processes in ultraperipheral collisions (UPCs) at the Large Hadron Collider (LHC) and the future Electron-Ion Collider (EIC).
Despite being fundamental in their own right and essential for QCD phenomenology of hard processes initiated by resolved photons, the quark and gluon distributions of the photon are poorly known and merit a new analysis employing modern tools of statistical data analysis. We determine new sets of leading-order (LO) and next-to-leading-order (NLO) parton distributions (PDFs) for the real photon, dubbed VALO1.0, by performing a global QCD analysis of the world data on the photon structure function $F_2^{\gamma}$ measured in deep-inelastic scattering (DIS) processes on a real photon target in electron-positron collisions. Our analysis improves on the results available in the literature by providing uncertainties in the form of Monte Carlo replicas and the photon PDFs in the LHAPDF6 format. We observe that while the electron-positron data allow us to determine the singlet quark distribution very well at both LO and NLO, the gluon distribution is constrained to a much lesser degree, especially at LO.
For this analysis, we have developed an open-source framework, which extends the \href{https://nnpdf.github.io/pineline/}{pineline} framework, developed for the proton,
to the photon case and includes the program solving the inhomogeneous scale evolution of photon PDFs, \href{https://geko.readthedocs.io/en/latest/}{$\gamma$EKO}.
}

\begin{document}
\maketitle
\flushbottom

\section{Introduction}

Despite being an elementary gauge boson, the photon can develop an effective hadronic structure through quantum fluctuations.
As a consequence,
%vgApart 
apart
from interacting directly with charged particles through its electromagnetic coupling, the photon may also participate in strong interactions via hadronic intermediate states.
%~\cite{Nisius:1999cv}.
This hadronic, the so-called resolved structure, arises either from the point-like splitting of the photon into a quark-antiquark pair, referred to as the anomalous (or point-like) contribution, or from its non-perturbative component, which traditionally has been described by the vector meson dominance (VMD) model.
%vg 
%\sout{in which the photon is treated as a superposition of vector mesons.}
Consequently, this implies that at high energies, a hard probe can resolve the underlying quark-gluon (parton) content of the photon, much like in the case of protons or nuclei, and it is possible to define parton distribution functions (PDFs) of the real photon $f^\gamma_j(j=q,g)$ to encode this structure.
This can be realized through electron-positron scattering, where 
one lepton emits a highly virtual photon that probes the quasi-real photon emitted by the other, resulting effectively in electron–photon deep inelastic scattering (DIS), $e + \gamma \xrightarrow[]{} e' + X$. 
We perform a global QCD analysis of the world data on the photon structure function $F_2^\gamma$ measured at CERN, DESY, KEK, and SLAC,
by implementing hadron-like initial conditions for the quark and gluon distributions and quantifying uncertainties via Monte Carlo replicas.
As a result, we determine new sets of LO and NLO photon
PDFs with uncertainties, which represent a key upgrade over existing analyses in the literature. We refer to them as VALO1.0 PDFs~\cite{Chithirasreemadam:2026mqp},
with the name coming from the Finnish word for ``light''.

%Photon PDFs can be further
%constrained through dijet photoproduction in electron–proton scattering measured at HERA, providing additional sensitivity to the gluon distribution. Ultraperipheral collisions at the Large Hadron Collider and photoproduction at the future Electron-Ion Collider may also serve as a probe and, more importantly, as a new application of photon PDFs.
%\textcolor{red}{\bf 1. Somewhat critically, why do we need this paragraph? Is there a nice way to connect it to the prevous text?
%2. Do we want to give a couple of key references for UPCs and EIC?}

% \section{Theory}

\section{The photon structure function and scale evolution of photon PDFs}

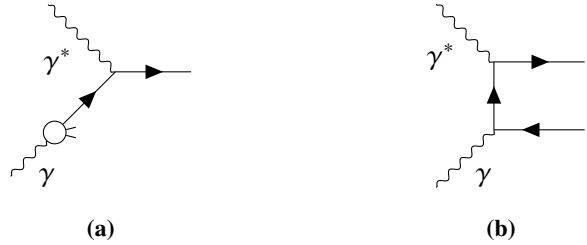
\begin{figure}[h]
\centering
\begin{subfigure}{0.35\textwidth}
\centering
\begin{tikzpicture}
   \begin{feynman}
        \vertex (mu);
        \vertex [right=2*0.5cm of mu] (f1);
        \vertex [below=0.5*0.7cm of f1] (f2);
        \vertex [above=0.5*0.7cm of f1] (f3);
        \vertex [above left=1.8*0.7cm of mu] (q) ;
        \vertex [below =1.5*0.7cm of mu] (alpha);
        \vertex [right=2*0.7 cm of alpha] (f2) ;
        \vertex [blob, minimum size = 0.3 cm, fill = none, below left= 1.4*0.7cm of mu](r){};
        \vertex [below left=0.9*0.7cm of mu] (l1);
        \vertex [right = 0.8*0.7cm of mu] (l2) ;
        \vertex [right = 0.4*0.7cm of r] (x) ;
        \vertex [above = 0.1*0.7cm of x] (r1) ;
        \vertex [below = 0.1*0.7cm of x]  (r2) ;
        \vertex [below left=2.8*0.7cm of mu] (p);
        \diagram* {
        (q) -- [photon, edge label'=\(\gamma^*\)] (mu) -- [fermion] (f1),
        (r) -- [fermion] (mu), 
        (p) -- [photon, edge label'=\(\gamma\)] (r),
        (r1) -- (r) -- (r2),
        };
    \end{feynman}
\end{tikzpicture} 
\caption{}
\label{subfig:resolved}
\end{subfigure}%
\begin{subfigure}{0.35\textwidth}
\centering
\begin{tikzpicture}
    \begin{feynman}
        \vertex (mu);
        \vertex [right=2*0.6cm of mu] (f1);
        \vertex [above left=1.8*0.6cm of mu] (q) ;
        \vertex [below =1.5*0.6cm of mu] (alpha);
        \vertex [right=2*0.6 cm of alpha] (f2) ;
        \vertex [below left=1.8*0.6cm of alpha] (p);
        \diagram* {
        (q) -- [photon, edge label'=\(\gamma^*\)] (mu) -- [fermion] (f1),
        (alpha) -- [fermion] (mu),
        (p) -- [photon, edge label'=\(\gamma\)](alpha),
        %c -- [anti fermion] f2 [particle=\(\overline \nu_{e}\)],
        (alpha) -- [anti fermion](f2),
        };
    \end{feynman}
\end{tikzpicture}
\caption{}
\label{subfig:boxed}
\end{subfigure}%
\caption{The LO (a) and the NLO point-like (b) contributions to the photon structure function $F_2^{\gamma}$. }
    \label{fig:LOFeynman}
\end{figure}

% \sout{The photon structure function $F_2^\gamma$ receives contributions from quarks at all orders, while gluons and the photon contribute at NLO, as shown in \cref{fig:LOFeynman}.}
As discussed in Introduction, the photon structure function $F_2^\gamma$ in electron-photon DIS 
receives two types of contributions: the one associated with the non-perturbative partonic content of the photon 
and the one proceeding thought the point-like coupling of the photon to electrically-charged quarks,
see graphs $a$ and $b$ in \cref{fig:LOFeynman}, respectively.
At LO, only the quarks contribute, as shown in graph $a$, while gluons and the term associated with
graph $b$ enter at NLO. As a result, at NLO accuracy and in the $\overline{\text{MS}}$ factorization scheme, the photon structure function $F_2^\gamma$ can be written as 
\begin{equation}
\frac{1}{x}F_2^{\gamma}(x,Q^2)=\qty(\vb C(\alpha_s(Q^2)) \otimes \vb f^{\gamma}(Q^2))(x)
+ \qty(n_u e_u^4 + n_d e_d^4) \frac{\alphaem}{4 \pi}C_{\gamma}(x) \,,
\label{eq:F2_MSbar_3}
\end{equation}
where we have used the compact vector notation for photon PDFs $\bf f^{\gamma}$, the coefficient functions $\bf C$ (in bold throughout this text),
and their convolution $\otimes$,
while $n_d$ and $n_u$ denote the numbers of up-type ($u$ and $c$) and down-type ($d, s$, and $b$) quarks
with the electric changes $e_u$ and $e_d$,
respectively,
and $\alphaem$ is the fine-structure constant.
% \sout{Unlike in the case of the proton, $F_2^\gamma$ exhibits positive scaling violations at all momentum fractions $x$. This is due to the point-like coupling of photons to quarks (\cref{subfig:boxed}), hidden in the photon coefficient function $C_\gamma$,}
%
The photon coefficient function $C_\gamma$ can be calculated exactly using the graph in \cref{subfig:boxed},
\begin{equation}
C_{\gamma}^{(1)}(x) = 4 N_c  \Bigg[\left(x^2+(1-x)^2\right) \ln\left(\frac{1-x}{x}\right)-1+8x(1-x) \Bigg] \,,
\label{eq:C1_gamma}    
\end{equation}
where $N_c=3$ is the number of colors. 
%It is the presence of this term in \cref{eq:F2_MSbar_3} 
%that drives positive scaling violations of $F_2^\gamma$ at all momentum fractions $x$.
To avoid numerical instabilities of $F_2^{\gamma}$
at large-$x$, it is convenient to work in the $\text{DIS}_\gamma$ factorization scheme~\cite{Gluck:1991ee}, where the point-like contribution is absorbed into the definition of the quark and antiquark PDFs. The scheme transformation is given by
%vg,
%\textcolor{red}{\bf Shall we try to convert it into the vector notation for uniformity and to save space?}
%\begin{align}
\begin{equation}
q_j^{\gamma}(x,Q^2)_{\rm DIS_{\gamma}} = q^{\gamma}_j(x,Q^2)_{\overline{\rm MS}}
+e_{q_j}^2\, \frac{\alphaem}{8 \pi} C_{\gamma}(x) \,.
%vg, \nonumber\\
%g^{\gamma}(x,Q^2)_{\rm DIS_{\gamma}} &= g^{\gamma}(x,Q^2)_{\overline{\rm MS}} \,.
\label{eq:scheme}
%\end{align}
\end{equation}

% \subsection{Scale evolution and Fitting Methodology}

The parton-parton splittings, along with the photon-quark splitting, $\gamma \xrightarrow[]{} q\overline{q}$, lead to scale evolution equations for the photon PDFs, which modify the familiar Dokshitzer-Gribov-Lipatov-Altarelli-Parisi (DGLAP) equations~\cite{Gluck:1983bh}.
In particular,
the point-like contribution
gives rise to an inhomogeneous term 
so that the evolution equations
can be written in the following matrix form in the $x$ space,
\begin{equation}
\mu_F^2 \frac{\dd \vb f^{\gamma}(x,\mu_F^2)}{\dd \mu_F^2}=\vb k(x,\alpha_s(\mu_F^2))+(\vb P(\alpha_s(\mu_F^2))\otimes \vb f^{\gamma}(\mu_F^2))(x) \,,
\label{eq:ev_matrix}
\end{equation}
where $\textbf{P}(\alpha_s(\mu_F^2)$ is the matrix of the parton-parton DGLAP splitting functions and $\textbf{k}(\alpha_s(\mu_F^2))$ is the point-like splitting function in vector notation. 

%vg new paragraph

We solve \cref{eq:ev_matrix} numerically by developing
the evolution package $\gamma\text{EKO}$ \cite{felix_hekhorn_2025_16032673}, which
is built upon the existing EKO (Evolution Kernel Operator)~\cite{Candido:2022tld} framework.
% \sout{accounts for the anomalous component, and is built upon the existing EKO (Evoution Kernel Operator)~\cite{Candido:2022tld} framework.}
The solution to the evolution of photon PDFs is then expressed in terms of a new photon evolution operator $\Tilde{\textbf{E}}^\gamma$, which generalizes the hadronic evolution operator  $\Tilde{\textbf{E}}$,
\begin{equation}
    \tilde{\mathbf f}^\gamma(\mu_F^2) = \tilde{\mathbf E}^\gamma(\mu_F^2 \leftarrow \mu_0^2)\left[\tilde{\mathbf f}^\gamma(\mu_0^2)\right]
     = \tilde {\mathbf E}(\mu_F^2 \leftarrow \mu_0^2) \tilde{\mathbf f}^\gamma(\mu_0^2) -\int\limits_{\mu_0^2}^{\mu_F^2}\! \frac{\diff \mu^2}{\mu^2} \tilde{\mathbf E}(\mu_F^2 \leftarrow \mu^2) \tilde{\mathbf k}(a_s(\mu^2)) \,, 
     \label{eq:gEKOop}
\end{equation}
where all the involved quantities with a tilde refer to Mellin moments of the functions entering \cref{eq:ev_matrix}, 
and $\tilde{\mathbf f}^\gamma(\mu_0^2)$ denotes the initial conditions for photon PDFs.

Further, the algorithm for quick computation of $F_2^\gamma$ relies on the Yadism package~\cite{Candido:2024rkr}, which allows one
to calculate the quark and gluon coefficient functions in \cref{eq:F2_MSbar_3}. We also borrow the fast-kernel (FK) formalism of the \texttt{pineline} framework
%~\cite{Barontini:2023vmr} 
by implementing hadronic and photonic FK tables, which are precomputed convolutions of the coefficient functions and the EKO (and $\gamma$EKO) operators.
It leads
%vgleading 
to a much simplified computation of $F_2^\gamma$ directly in terms of the initial condition $f^{\gamma}(Q_0^2)$, 
\begin{equation}
    \frac 1 x F_2^{\gamma}(x,Q^2) = \qty(\vb {FK}(Q^2,Q_0^2) \otimes \vb f^{\gamma}(Q_0^2))(x) + \mathrm{FK}^{\gamma}(x,Q^2) \,.
    \label{eq:factFK}
\end{equation}

%vg \section{Initial conditions and experimental datasets }
\section{Initial conditions, experimental data sets, and fit quality}
%\textcolor{red}{\bf I suggest to move the discussion of MC replicas after the initial condition}

\begin{figure}[t]
    \centering
    \includegraphics[width=0.6\linewidth]{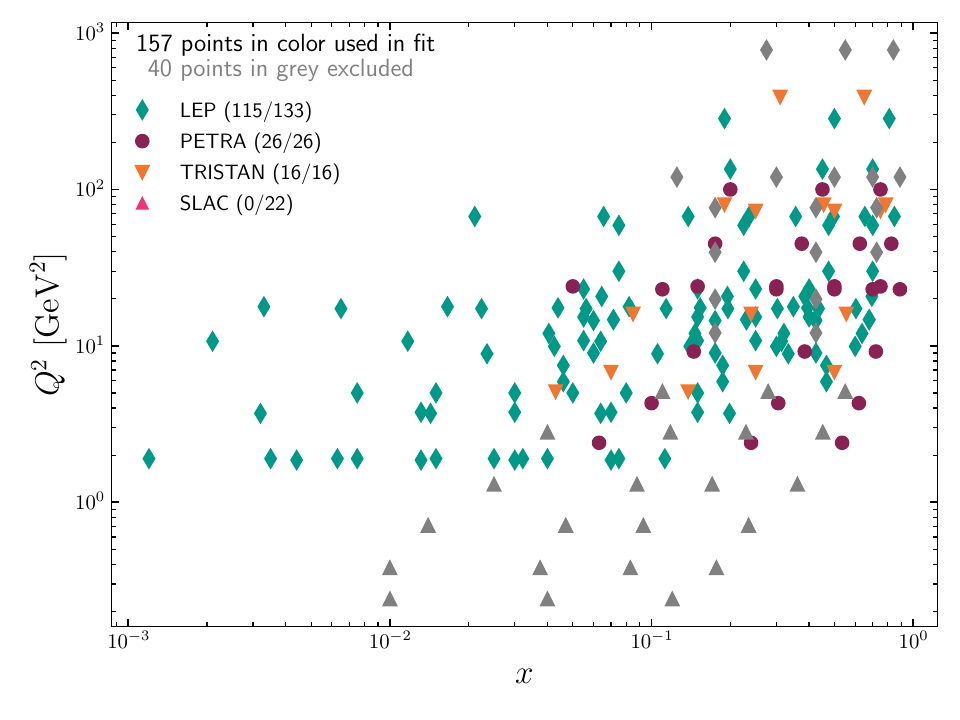}
    %vg\caption{Available photon structure function data points from $e^+e^-$ collisions used for our fits plotted in the $x, Q^2$ plane.Excluded data is plotted in gray.}
    \caption{The kinematic $x\text{-}Q^2$ coverage of the world data on the photon structure function
    $F_2^{\gamma}$ from $e^+e^-$ collisions. The colored symbols  (diamonds, circles, inverted triangles) show the data points used in our global fit, while the data in grey (diamonds, triangles) are excluded from our analysis. The data are labeled by the collider name, indicating also the number of used and total available points
    in parentheses.
    }
    \label{fig:kin}
\end{figure}

%\sout{We rely on the Monte-Carlo (MC) replicas framework for estimating PDF uncertainties in our global analysis, and present 100 replicas along with the central PDF in the LHAPDF format.}
The initial conditions of the photon PDF, $f_0^\gamma(x)\equiv f^\gamma(x,Q_0^2=\SI{1}{\GeV^2})$ are parameterized in the following hadron-like form
\begin{align}
\frac{1}{\alphaem} xu_0^\gamma(x) = \frac{1}{\alphaem} x\bar u_0^\gamma(x) &= N_u x^{a_u}(1-x)^{b_u}  \,,  \nonumber\\  
d_0^\gamma(x)  =   \bar d_0^\gamma(x) &=  u_0^\gamma(x) , \nonumber\\ 
    s_0^\gamma(x) = \bar s_0^\gamma(x) &=K_s \, u_0^\gamma(x) \,,  \nonumber\\
  \frac{1}{\alphaem}  xg_0^\gamma(x) &= N_gx^{a_g}(1-x)^{b_g} \,,
    \label{eq:input_param}
\end{align}
where we assume exact isospin symmetry and
vanishing valence quark distributions.
The parameters $N_u, a_u,b_u,N_g,b_g$ are five free parameters of our fit, while the large-$x$ exponent $b_g=3$ and  the strange suppression factor $K_s=0.3$
%vg, 
are fixed, motivated 
by the VMD model.
Note that we do not impose any momentum sum rule for the photon PDFs in our analysis. 

We rely on the Monte-Carlo (MC) replicas framework for estimating PDF uncertainties in our global analysis, and present 100 replicas along with the central PDF in the LHAPDF format.
Each replica is drawn from a multi-gaussian distribution centered around the measured data points and with its width (standard deviation) given by the associated experimental uncertainties.
%\textcolor{red}{\bf Madhav, please streamline this paragraph by adding the phrase that we treat experimental errors also using MC replicas. This way thewe will be a logical connection between your sentences describing MC replicas and the datasets. }
In \cref{fig:kin}, we show all 157 $F_2^\gamma$ data points available in the literature, where we have imposed a $Q^2\geq \SI{1}{\GeV^2}$ cutoff on the data used in these fits.

Our global QCD analysis converges well to the data on $F_2^\gamma$, yielding total $\chi^2$ per degree of freedom (DOF) of 0.81 and 0.94 at LO and NLO, respectively. As mentioned earlier, using $F_2^\gamma$ data, one can only probe the gluon content in a photon from NLO and therefore one could attribute the slight increase in $\chi^2/\text{DOF}$ at NLO to the rigid input-scale parameterization of the gluon distribution.

\section{Results}

\begin{figure}[t!]
\begin{center}
    \includegraphics[width=0.4\linewidth]{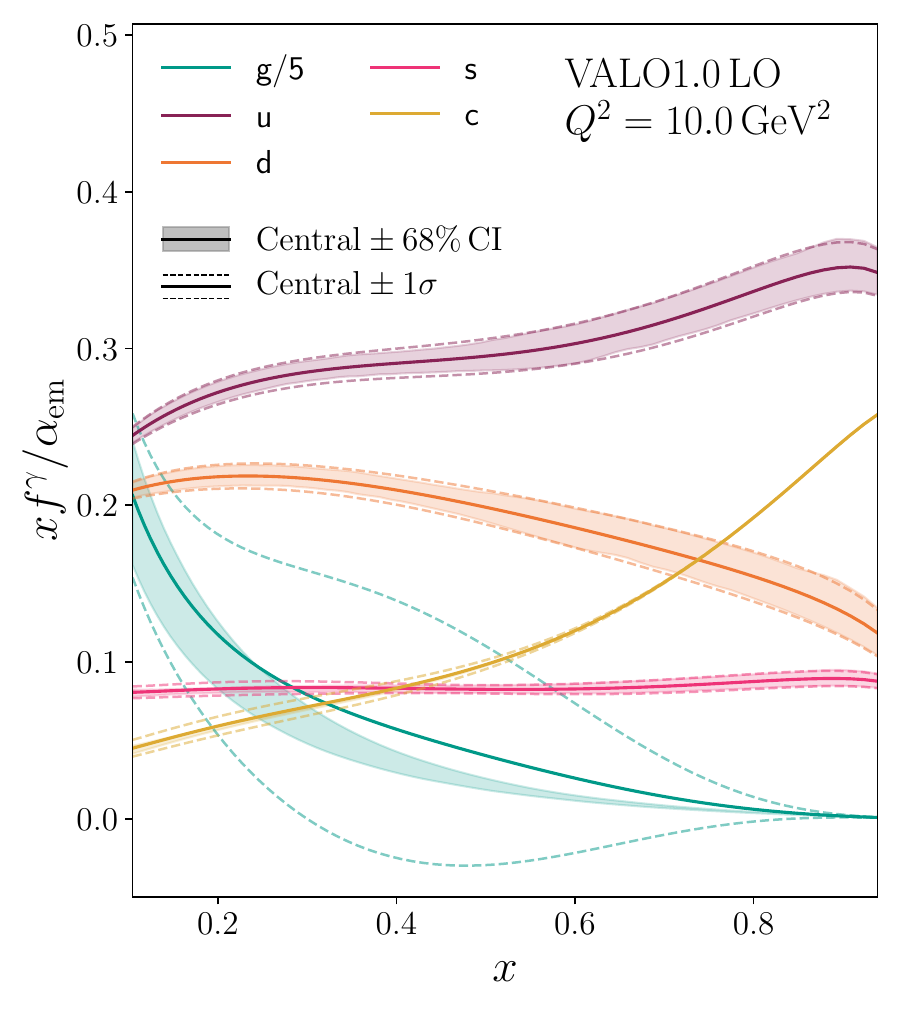}%
    \includegraphics[width=0.4\linewidth]{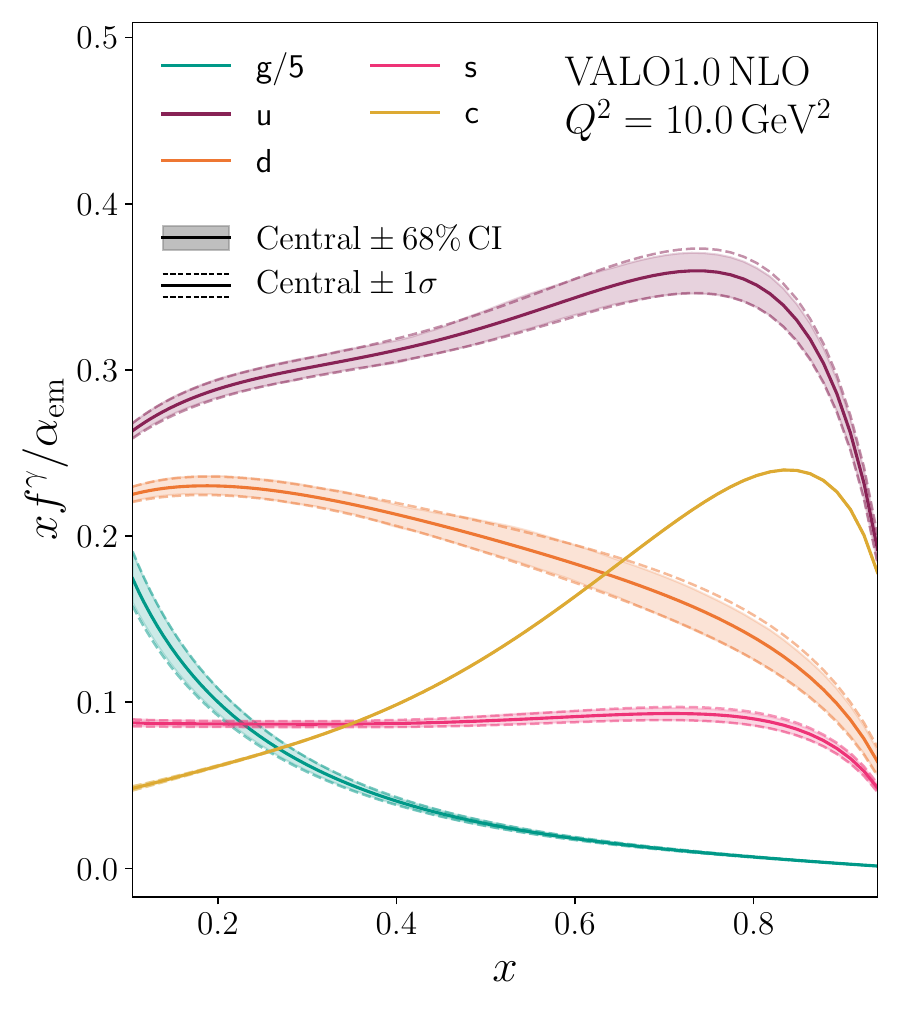}
    \end{center}
    %vg\caption{PDFs at $Q^2 = \SI{100}{\GeV^2}$ at LO (left) and NLO (right)}
    \caption{%vgEvolved 
    VALO1.0 photon PDFs for individual parton flavors $xf_j^{\gamma}(x,Q^2)/\alpha_{\rm em}$ as a function of $x$ at $Q^2=10$ GeV$^2$: the central PDFs and the uncertainty bands
    (standard deviation and 68\% CI) at LO (left panel) and NLO
    (right panel).
    }
    \label{fig:LONLOhigh}
\end{figure}

\begin{figure}[t!]
\begin{center}
    \includegraphics[scale=0.4]{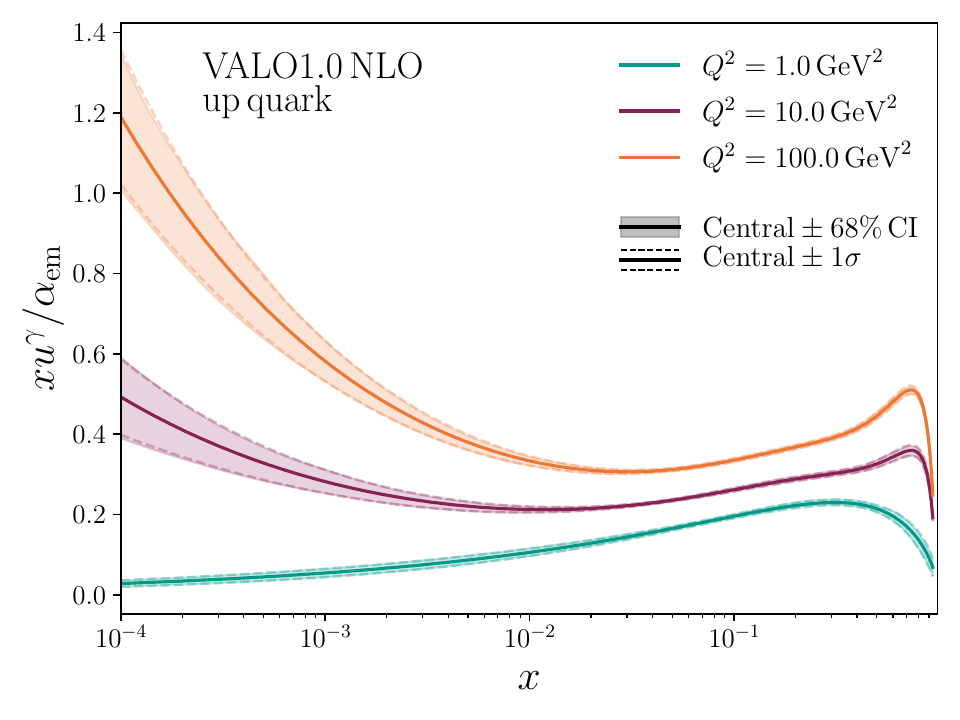}%
    \includegraphics[scale=0.4]{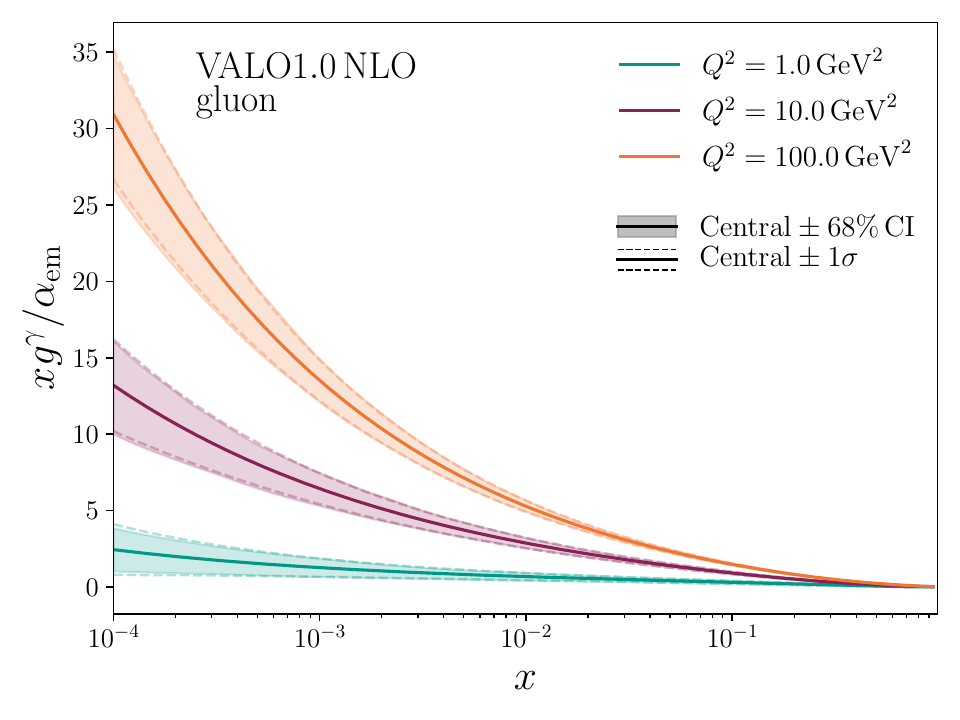}
    \end{center}
    %vg\caption{PDFs at $Q^2 = \SI{100}{\GeV^2}$ at LO (left) and NLO (right)}
    \caption{Small-$x$ behavior of NLO VALO1.0 photon PDFs
    %the LO (upper panels) and NLO
    %vg(lower panels) 
    for up-quark (left panel) and gluon (right panels) distributions as a function of $x$ at $Q^2=(1,10,100)$ GeV$^2$.}
    \label{fig:LONLOhighlog}
\end{figure}
\begin{figure}[t!]
    \centering
    \includegraphics[scale=0.45]{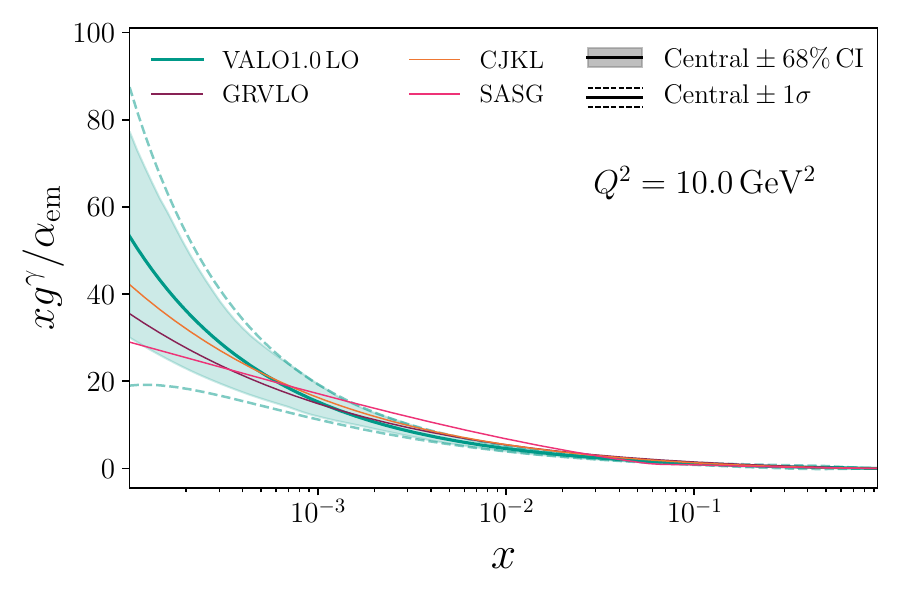}%    
    \includegraphics[scale=0.45]{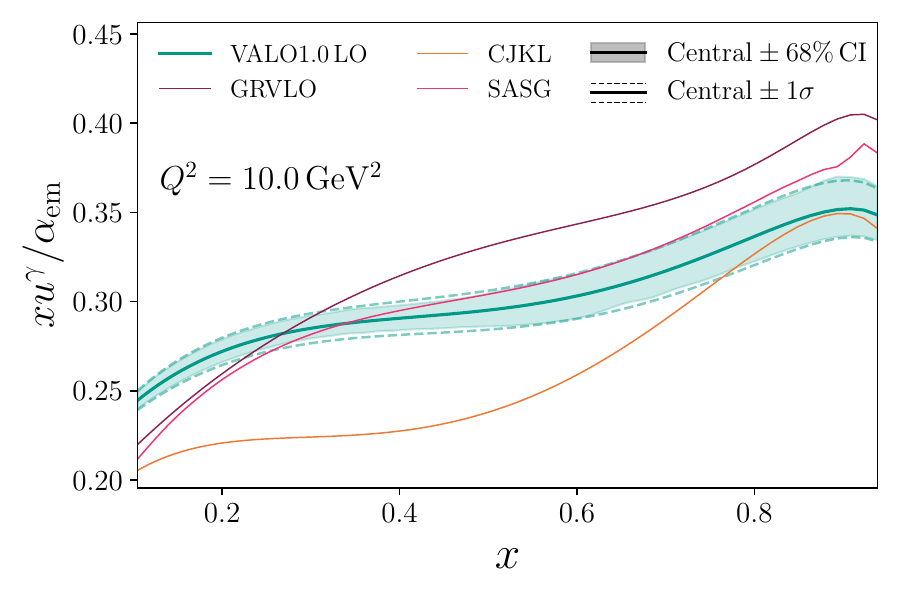}
    \includegraphics[scale=0.45]{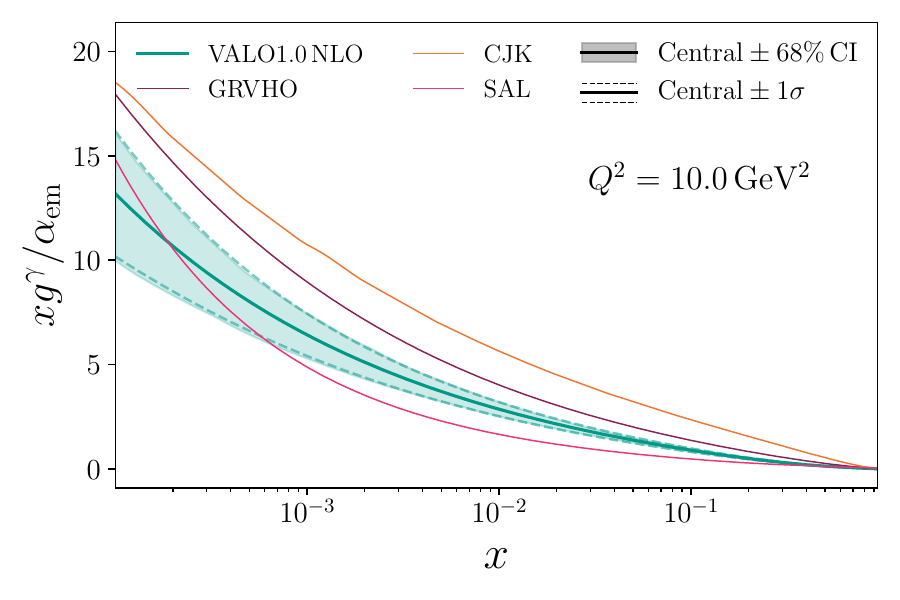}%    
    \includegraphics[scale=0.45]{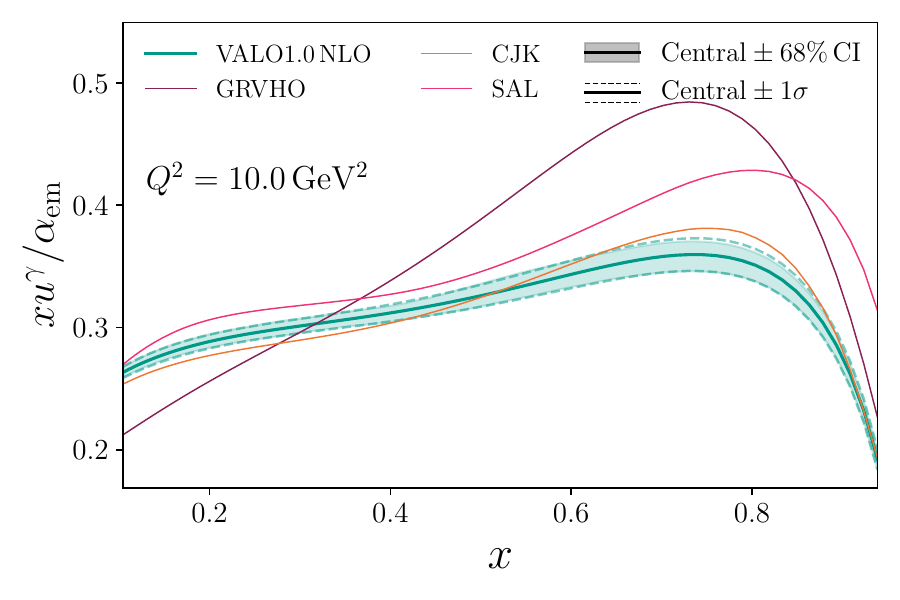}
    %vg\caption{Same as in \cref{fig:referancePDFsLO}, but comparison with existing NLO Photon PDFs GRVHO \cite{Gluck:1991jc}, CJK \cite{Cornet:2004nb}, SAL \cite{Slominski:2005bw} at $Q^2 = \SI{10}{\GeV^2}$.}
    \caption{Comparison of LO (upper panels) and NLO (lower panels) VALO1.0 PDFs 
    %vg(upper panels) photon PDFs 
    with the photon PDFs available in the literature 
    %GRVLO~\cite{Gluck:1991jc}, CJKL~\cite{Cornet:2002iy}, and  SASG~\cite{Schuler:1995fk} PDFs (upper panels) and  the NLO VA (bottom panels) photon PDFs with GRVHO~\cite{Gluck:1991jc},
    %CJK~\cite{Cornet:2004nb}, and SAL~\cite{Slominski:2005bw} PDFs 
    as a function of $x$ 
    at $Q^2=10$ GeV$^2$.
    %gluon PDFs at logarithmic scale (left), while up-quark PDFs in linear scale (right).
    }
    \label{fig:referancePDFsNLO}
\end{figure}

% \sout{We present the VALO1.0 photon PDFs in this section in the form of raw PDFs, as well as through comparisons with several existing PDF sets.}
% \sout{Our global QCD analysis converges very well to the data on $F_2^\gamma$, yielding total $\chi^2$ per degree of freedom (DOF) of 0.81 and 0.94 at LO and NLO, respectively. As mentioned earlier, using $F_2^\gamma$ data, one can only probe the gluon content in a real photon from NLO. We could therefore attribute the slight increase in $\chi^2/\text{DOF}$ at NLO to the rigid input-scale parameterization of the gluon distribution.}

%\sout{We present our photon PDFs as $xf_j^\gamma(x,Q^2)/\alpha_{\text{em}}$ in subsequent figures.} 
In \cref{fig:LONLOhigh}, we show 
%vgscale evolved 
the LO (left panel) and NLO (right panel) VALO1.0 photon PDFs 
%vgat $Q^2=\SI{10}{\GeV^2}$ 
for individual parton flavors as a function of the momentum fraction
$x$ at $Q^2=\SI{10}{\GeV^2}$.
%
%, as a function of momentum fraction
%vg, 
%vg, at LO (left) and NLO.
While the quark PDFs are rather robust, 
%vgacross all orders, 
the gluon PDF at LO is not well constrained 
%vg, as can be seen in \cref{fig:LONLOhigh} with 
since 
the standard deviation (dotted lines) and the \SI{68}{\percent} confidence interval (uncertainty bands) 
are 
not in agreement. The systematic and statistical errors in the experimental data are small, 
%vgas reflected in our PDFs by the narrow uncertainty bands for quarks.
which is reflected in narrow uncertainty bands of the resulting quark PDFs.
In contrast to the proton case, %vgpoint-like contribution to the 
the inhomogeneous term in the scale evolution 
leads to non-vanishing quark PDFs at the $x\xrightarrow[]{}1$ limit after evolution.
%vgalthough they rapidly fall at very large $x$ at NLO.
The low-$x$ behaviour of the NLO up-quark and gluon PDFs is shown in the left and right panels of \cref{fig:LONLOhighlog},
respectively.
%, at NLO. 
One can see from the figure that 
the quark and gluon PDFs steeply rise at small $x$,
%vgThe gluon PDFs see a steep rise at small $x$, 
just like in the case of proton PDFs. At the same time, 
%vgwhile 
the uncertainty bands of both quark and gluon PDFs become wider, indicating a lack of experimental data coverage in the region.
One should also note the positive scaling violations for all $x$ and the large-$x$ peak of the up-quark distribution, which are both driven by the inhomogeneous term in the scale evolution of photon PDFs originating from the $\gamma^{\ast} \to q {\bar q}$ point-like coupling. 

Finally, in~\cref{fig:referancePDFsNLO}, we compare our PDFs to existing sets in the literature: GRVLO~\cite{Gluck:1991jc}, CJKL~\cite{Cornet:2002iy}, and SaSG~\cite{Schuler:1995fk} at LO; GRVHO~\cite{Gluck:1991jc}, CJK~\cite{Cornet:2004nb}, and SAL~\cite{Slominski:2005bw} at NLO.
We show the gluon PDFs at logarithmic scales, while the up-quark distributions are plotted linearly with $x$. We agree well at both orders, while the biggest numerical differences are in the small-$x$ limit, where the uncertainty bands reasonably cover the differences.

\section{Conclusion}
We perform a global QCD analysis on $F_2^\gamma$ data from $e^+e^-$ scattering and extract new VALO1.0 PDFs of the real photon
with uncertainties quantified using MC replicas.
%The VALO1.0 PDFs offer a consistent description of the photon hadronic structure, with quantified MC replicas to estimate uncertainties.
We provide the photon PDFs at LO and NLO in both $\text{DIS}_\gamma$ and $\overline{\text{MS}}$ factorization schemes~\cite{chithirasreemadam_2026_19709212}, our fitting framework \texttt{VALOfitter} \cite{hekhorn_2026_19694026}, and the photon evolution code \texttt{gEKO} \cite{felix_hekhorn_2025_16032673}.
Photon PDFs can be further
constrained through dijet photoproduction in electron–proton scattering measured at HERA, providing additional sensitivity to the gluon distribution. Ultraperipheral collisions at the Large Hadron Collider and photoproduction at the future Electron-Ion Collider may also serve as a probe and, more importantly, as new applications of photon PDFs.

\acknowledgments
This research has been funded by the Center of Excellence in Quark Matter of the Research Council of Finland, projects 364191, 364194, and the Research Council of Finland project 361179. This research is associated with the Doctoral Education Pilot initiative for nuclear and particle physics of the Ministry of Education and Culture.

% Bibliography

\bibliography{biblio}{}
\bibliographystyle{JHEP}

\end{document}